\documentclass[useAMS,usenatbib]{mn2e}
\usepackage{graphicx}
\usepackage{times}

\title[Dust constraints at $z\sim9.6$]{Early Science with the Large Millimeter Telescope: 
Dust constraints in a $\bmath{z\sim9.6}$ galaxy}
\author[J. A. Zavala et al.]{J. A. Zavala\thanks{E-mail: zavala@inaoep.mx}$^{1}$, M. J. Micha{\l}owski$^{2}$, I. Aretxaga$^{1}$, G. W. Wilson$^{3}$,  D. H. Hughes$^{1}$,  
\newauthor A. Monta\~na$^{1,4}$,  J. S. Dunlop$^{2}$,  A. Pope$^{3}$, D. S\'anchez-Arg\"uelles$^{1}$, M. S. Yun$^{3}$
\newauthor  and M. Zeballos$^{1}$\\
$^{1}$Instituto Nacional de Astrof\'{i}sica, \'{O}ptica y Electr\'{o}nica (INAOE),
Luis Enrique Erro 1, Sta. Ma. Tonantzintla, Puebla, Mexico\\
$^{2}$Institute for Astronomy, University of Edinburgh, Royal Observatory, Blackford Hill, Edinburgh, EH9 3HJ, UK\\
$^{3}$Department of Astronomy, University of Massachusetts, MA 01003, USA\\
$^{4}$Consejo Nacional de Ciencia y Tecnolog\'ia (CONACyT), Av. Insurgentes Sur 1582, 03940, D.F., Mexico}
\begin{document}
\date{Accepted 2015 July 21.  Received 2015 July 20; in original form 2015 June 26.}
%

\maketitle
\label{firstpage}

\begin{abstract}
Recent observations with the GISMO (Goddard-IRAM Superconducting 2 Millimeter Observer) 
2 mm camera revealed a detection $8$ arcsec away from
the lensed galaxy MACS1149-JD1 at $z=9.6$. Within the $17.5$ arcsec FWHM GISMO beam, this
detection is consistent with the position of the high-redshift galaxy and therefore,
if confirmed, this object could be claimed to be the youngest galaxy producing significant quantities of  dust. 
We present higher resolution ($8.5$ arcsec) observations of this system taken with the AzTEC 1.1 mm
camera mounted on the Large Millimeter Telescope {\it Alfonso Serrano}. Dust continuum
emission at the position of MACS1149-JD1 is not detected with an r.m.s. of 0.17 mJy/beam.
However, we find a detection $\sim11$ arcsec away from MACS1149-JD1, still within 
the GISMO beam which is consistent with an association to the GISMO source. Combining the 
AzTEC and GISMO photometry, together with {\it Herschel} ancillary data, we derive a 
$z_{\rm phot}= 0.7-1.6$ for the dusty galaxy. We conclude therefore that the GISMO and AzTEC detections are not 
associated with MACS1149-JD1. From the non-detection of MACS1149-JD1  
we derive the following ($3\sigma$) upper limits corrected for 
gravitational lensing magnification and for cosmic microwave background (CMB) effects: dust mass $<1.6\times10^7$ M$_\odot$,
IR luminosity $<8\times10^{10}$ L$_\odot$, star formation rate $<14$ M$_\odot$ yr$^{-1}$,
and UV attenuation $<2.7$ mag. These limits are comparable
to those derived for other high-redshift galaxies from deep Atacama Large Millimeter/submillimeter Array (ALMA) observations.
\end{abstract}


\section{Introduction}
Dust is an extremely important ingredient of the Universe and plays an essential role 
in the formation and growth of different astronomical objects, from planets to the most 
massive galaxies. Dust obscures  ultraviolet (UV) and optical emission from young stars 
in regions of intense star-formation and re-emits it into far-infrared (FIR) bands, 
affecting the observed spectral properties (e.g. \citealt{2003ARA&A..41..241D}). 
Understanding the properties of dust in the very young universe is of great importance, 
since its abundance and composition provide critical information about the interstellar 
medium conditions at the epoch of reionisation, as well as about the number and total
energy output of stellar sources which likely contributed to the process of reionisation.
Furthermore, observations at these early epochs, when dust was being formed for the first 
time, have important consequences on our understanding of dust formation processes and 
the chemical enrichment of the Universe (e.g. \citealt{2007ApJ...662..927D}; 
\citealt{2010A&A...522A..15M}; \citealt{2014MNRAS.438.2765C}; \citealt{2014MNRAS.444.2442V};
\citealt{2015arXiv150501841M}; \citealt{2015arXiv150308210M}).

Dust has been detected in the $z=6.3$ submm galaxy HLFS3 (\citealt{2013Natur.496..329R}), 
the quasars SDSS J1148+5251 at z = 6.42 (\citealt{2003AJ....125.1649F}) and ULAS J1120+0641 
at $z=7.1$ (\citealt{2012ApJ...751L..25V}), and the $z=7.5$ 'normal' star-forming galaxy 
A1689-zD1 (\citealt{2015Natur.519..327W}). On the other hand,  recent interferometric studies
report non-detections of dust in $z \sim 6$--8 galaxies (\citealt{2012ApJ...752...93W}; 
\citealt{2013ApJ...778..102O}; \citealt{2014ApJ...796...96B}; \citealt{2014ApJ...784...99G}; 
\citealt{2014ApJ...792...34O}; \citealt{2015A&A...574A..19S}), leading to the belief that most 
normal galaxies at these high-redshifts may be largely dust-free due to their presumed low
metallicities. This is supported by the demonstration of a very low mass dust reservoir
in IZw18, a metal-poor local galaxy (\citealt{2014Natur.505..186F}; \citealt{2014A&A...561A..49H}). 
Clearly, the presence and abundance of dust in the high-redshift Universe is still poorly 
understood and, therefore, observations of dust of larger samples of galaxies at high-redshifts 
are necessary. 

MACS1149-JD1 is believed to be a gravitationally lensed galaxy, which is strongly magnified ($\mu=14.5^{+4.2}_{-1.0}$, 
\citealt{2012Natur.489..406Z}) by the Frontier Field (FF) galaxy cluster MACS J1149.6+2223 at $z=0.544$ (\citealt{2007ApJ...661L..33E}). The 
{\it Hubble Space Telescope (HST)} and {\it Spitzer} colours indicate that MACS1149-JD1 is a young  galaxy with a
photometric redshift of $z_{\rm phot}=9.6\pm0.2$ (\citealt{2012Natur.489..406Z}). The large number of 
bands used to derive the photometric redshift makes it one of the most accurate estimates obtained 
for such a distant object. Recently, \citet{2014ApJ...788L..30D} have  reported a potential 2 mm 
detection of MACS1149-JD1, using the Goddard-IRAM Superconducting 2 Millimeter Observer (GISMO) 
camera (\citealt{2008SPIE.7020E..04S}) mounted at the Institut de Radioastronomie Millim\'etrique 
(IRAM) 30m telescope, with a $17.5$ arcsec full width at half maximum (FWHM) beam size, but with an 
effective $24.7$ arcsec FWHM smoothed-beam for point-source extraction. The 99 per cent confidence 
interval for the positional uncertainty of this source is $\sim11$ arcsec, following the description by
\citet{2007MNRAS.380..199I}, consistent with the offset from the high-redshift galaxy ($\sim8$ arcsec). 
Furthermore, the 2 mm number counts (\citealt{2014ApJ...790...77S}) predicts a probability of 
0.3--0.8 per cent to find such a source at random, which suggests that the GISMO detection might be 
associated with the  high-redshift galaxy. 

In this Letter, we present $8.5$ arcsec FWHM resolution observations taken during the Early Science 
Phase of the Large Millimeter Telescope {\it Alfonso Serrano} (LMT; \citealt{2010SPIE.7733E..12H}) 
using the 1.1 mm continuum camera AzTEC (\citealt{2008MNRAS.386..807W}) in order to confirm or rule
out the association of the mm-wavelength source with MACSJ1149-JD1, as well as constrain the dust properties
of this system.

All our calculations assume a $\Lambda$ cold dark matter cosmology with $\Omega_\Lambda=0.68$, $\Omega_{\rm m}=0.32$, and 
$H_0=67$ kms$^{-1}$Mpc$^{-1}$ (\citealt{2013arXiv1303.5076P}).

\section{OBSERVATIONS}
\subsection{AzTEC/LMT observations}

Our observations were obtained using the 1.1 mm continuum camera AzTEC on the LMT. During this 
Early Science Phase (see also \citealt{2015MNRAS.452.1140Z}), only the inner 32 m diameter section of the LMT primary surface is illuminated,
leading to an effective beam size of $\theta_{\rm FWHM}=$8.5 arcsec, a factor of $\sim3$ better than the 
smoothed-GISMO/IRAM beam.

The observations were conducted in photometry mode over several observing nights between 2014 December and 2015 April with
opacities in the range of $\tau_{\rm 225GHz}= 0.03$--0.12. The scanning technique used was  a 
Lissajous pattern covering a 1.5 arcmin diameter region with uniform noise centred at the {\it HST}
position of MACS1149-JD1. Pointing observations towards the bright millimetre source J1159+292 
($S_{1{\rm mm}}\approx$2 Jy) were acquired every 30 min to ensure positional accuracy. A total on-source 
integration time of $\sim5$ h was obtained in the photometry mode map, achieving an r.m.s. of 0.19 mJy. Additional data from  
a wider area  ($\sim16$ arcmin$^2$) map of the MACS cluster (Monta\~na et al. in preparation; Pope et al. in preparation) were co-added to
obtain a final r.m.s. of 0.17 mJy at the position of MACS1149-JD1. These observations have a similar depth 
to the GISMO observations at 2 mm assuming  an average SMG spectral energy distribution (SED) template 
(\citealt{2010A&A...514A..67M}) at $z\sim9.6$. Alternatively, the map is $\sim4$ times deeper if the 
SMG template is located at a less extreme redshift of $\sim1$. To generate the final 1.1 mm maps from the raw data we used the AzTEC 
Standard Pipeline described by \citet{2008MNRAS.385.2225S}.

The distance between 15 AzTEC sources with S/N$>4$ detected in the full  map and the nearest 
{\it Spitzer} Infrared Array Camera 4.5 $\mu$m sources has a 1$\sigma$ offset of $2.25$ arcsec, including the overall systematic
offset between the two images and the positional uncertainty in AzTEC source positions. Therefore, 
the astrometric accuracy of the AzTEC image is estimated to be better than $2.25$ arcsec.

\subsection{Ancillary  data}\label{ancillary}
We use public {\it Herschel} Lensing Survey  (HLS;\footnote{http://herschel.as.arizona.edu/hls/hls.html}
\citealt{2010A&A...518L..12E}) data in order to complete the SED of our
AzTEC detection (see Section \ref{aztec_detection}). HLS is an imaging survey of massive galaxy clusters
in the FIR and submillimetre using the {\it Herschel Space Observatory} (\citealt{2010A&A...518L...1P}), that includes 
observations from both PACS (\citealt{2010A&A...518L...2P}) and SPIRE (\citealt{2010A&A...518L...3G})
instruments at 100, 160, 250, 350, and 500 $\mu$m, with angular resolutions of $\sim8$, 13, 18, 25, 36 arcsec, 
respectively.

\begin{figure}
\begin{center}
\includegraphics[width=90mm]{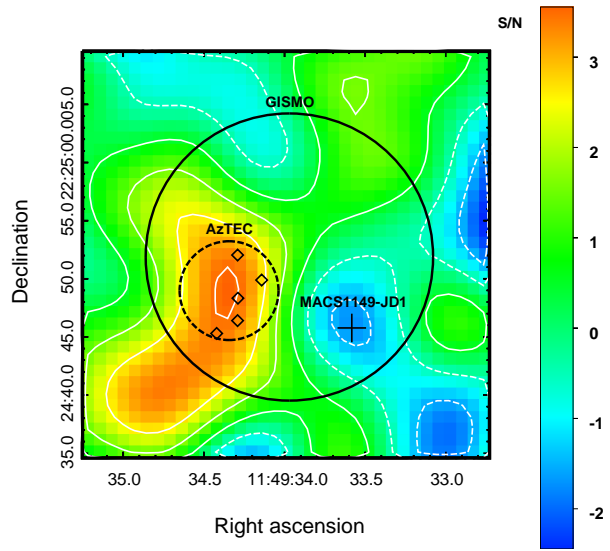}
\caption{AzTEC 1.1 mm signal-to-noise map around the MACS1149-JD1 field. The white contours
denote $-1.5\sigma$, $-0.5\sigma$ (dashed), and $+0.5\sigma$, $+1.5\sigma$,$+2.5\sigma$, 
and $+3.5\sigma$ (solid) significance levels. The 3.5$\sigma$ AzTEC detection (black dashed $8.5$ arcsec 
diameter circle) is $11$ arcsec  away from the MACS1149-JD1 position (black cross), and inside the smoothed
GISMO beam (24.7 arcsec FWHM, black solid circle). The black diamonds indicate the position of  
potential counterparts to the (sub-)mm source from the CLASH catalogue at $z\sim0.7-1.6$, consistent 
with the photometric redshift of our AzTEC source (see Fig. \ref{phot_z}). No 1.1 mm
continuum emission is detected at the position of MACS1149-JD1.} 
\label{mapaSN}
\end{center}
\end{figure}

To search for possible optical counterparts to our AzTEC detection 
we also use a catalogue from the Cluster Lensing And Supernovae survey with {\it Hubble} 
(CLASH, \citealt{2012ApJS..199...25P}), which consists of {\it HST} observations 
of massive galaxy clusters. CLASH observations are carried out in 16 bands from  the UV to the NIR
($0.2-1.7$ $\mu$m) and can be used to derive reliable estimates of optical-IR photometric redshifts.

\begin{figure*}
\includegraphics[width=170mm]{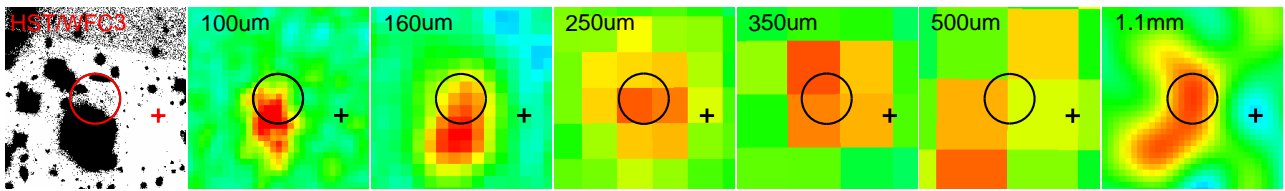}
\caption{$30$ arcsec $\times 30$ arcsec postage stamps with {\it HST} {\it f160w} band, {\it Herschel} 
$100-500$ $\mu$m bands, and AzTEC 1.1 mm centred at the position of the AzTEC-detected source. The circle 
represents the $8.5$ arcsec FWHM beam and position of the AzTEC source. The position of MACS1149-JD1 is marked 
with a  cross. The AzTEC source is likely related to a blend of the galaxies within 
the circle in the {\it HST} map.}
\label{postages}
\end{figure*}

\begin{table*}
\begin{minipage}{126mm}
\caption{Photometry measurements of the (sub-)mm-detected source. }
\begin{tabular}{ccccccc}
\hline
$S_{100\mu m}$& $S_{160\mu m}$&$S_{250\mu m}$&$S_{350\mu m}$&$S_{500\mu m}$&$S_{1.1mm}$&$S_{2mm}$\\
(mJy)         &(mJy)         &(mJy)         &(mJy)         &(mJy)          &(mJy)          &(mJy)\\
\hline
$6.3\pm0.3$&$13.3\pm0.7$&$16.0\pm2.0$&$9.8\pm2.8$&$8.3\pm3.3^a$&$0.62\pm0.17$&$0.40\pm0.10$   \\

\hline
\multicolumn{7}{l}{$^a$We consider the $2.5\sigma$ detection to be an upper limit.}
\label{photometry}
\end{tabular}
\end{minipage}
\end{table*}

\section{Results}
\subsection{The (sub-)millimetre source around MACS1149-JD1}\label{aztec_detection}

Despite achieving 0.17 mJy r.m.s. we have not found evidence of 1.1 mm
continuum emission at the position of MACS1149-JD1. The closest source in our 1.1 mm map has a $11$ arcsec offset
from MACS1149-JD1 (see Fig. \ref{mapaSN}), and is detected with an S/N$=3.5$ 
($S_{1.1mm}=0.62\pm0.17$ mJy) at  RA$=11^h49^m34^s.35$ Dec$=+22^\circ24'49''.1$.

In order to measure the positional uncertainty of this AzTEC source, we randomly insert and extract 
100 000 simulated sources from our AzTEC map following the procedure described in 
\citet{2010MNRAS.405.2260S}. These simulations use the actual AzTEC map and so include random and 
confusion noise and any associated biases present in the image.  We find no instance of a source located 
at a distance $\ge 11$ arcsec from the input position, and hence we infer that the probability that the AzTEC source is the counterpart of
MACS1149-JD1 is $\ll 10^{-4}$, and therefore we can reject the association. 
However, our detection lies inside of the GISMO beam (see Fig. \ref{mapaSN}), and hence both detections likely 
correspond to the same object. Furthermore, there are significant detections in some {\it Herschel} 
bands within the AzTEC beam (see Fig. \ref{postages}), which supports our association.

\begin{figure}
\begin{center}
\includegraphics[width=90mm]{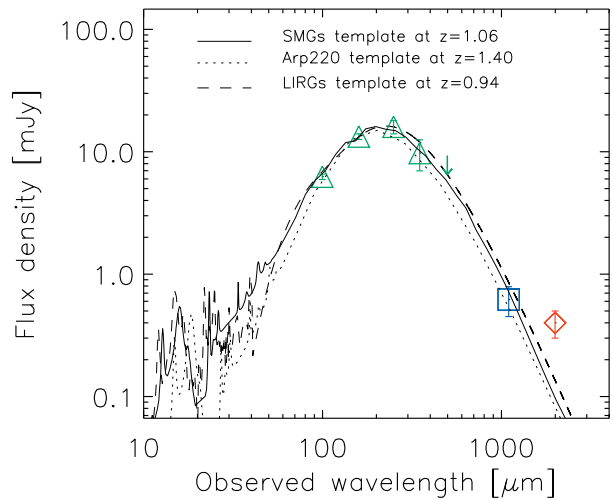}
\caption{Best-fitting SEDs for the (sub-)mm source using the Arp220, average SMG, and average LIRG 
distributions (see Section \ref{aztec_detection}). The photometry includes {\it Herschel} (green triangles),
AzTEC (blue square), and GISMO (red diamond) data (see Table \ref{photometry}). The orange diamond 
represents the GISMO deboosted flux density. The best fits give us a range of $z_{\rm phot}=0.7-1.6$.} 
\label{phot_z}
\end{center}
\end{figure}

We have obtained the {\it Herschel} fluxes at the AzTEC position by fitting a Gaussian function with
the corresponding beam FWHM of each {\it Herschel} map. We fit the  {\it Herschel}, AzTEC, and GISMO photometry (see Table \ref{photometry})
with different SED templates, including the local ULIRG galaxy Arp220 (\citealt{1998ApJ...509..103S}),
an average SMG template (\citealt{2010A&A...514A..67M}) and an average 24 $\mu$m-selected star-forming galaxy 
template, corresponding to LIRG luminosity (\citealt{2012ApJ...759..139K}). The range for the photometric redshifts obtained from the SED fitting is
$z_{\rm phot}=0.7-1.6$ (see Fig. \ref{phot_z}). This implies that this galaxy likely lies within the lower tail of the 
redshift distribution of SMGs (e.g. \citealt{2005ApJ...622..772C}; \citealt{2007MNRAS.379.1571A}; 
\citealt{2012MNRAS.426.1845M}; \citealt{2012MNRAS.420..957Y}; 
\citealt{2014MNRAS.443.2384Z}), although with an intrinsic luminosity and star formation rate (SFR) lower than the typical 
values found for this population  of galaxies. From the SED fitting we  derive an 
$L_{\rm FIR}=3\times10^{11}-1\times10^{12}$ $L_\odot$ $\mu^{-1}$ and SFR$=50-170$ $M_\odot$ yr$^{-1}$ $\mu^{-1}$ , 
where $\mu$ is the lens amplification which, based on public FF lensing 
models\footnote{www.stsci.edu/hst/campaigns/frontier-fields/Lensing-Models} (e.g. \citealt{2014MNRAS.444..268R}), 
is expected to be $\mu=1.2-7.4$ (depending on redshift and the adopted model).

Our fits do not reproduce well the 2 mm data point. However, confusion noise, which has been well 
characterized in a shallower GISMO 2 mm-map (\citealt{2014ApJ...790...77S}), may contaminate the 
measured flux density, as well as surrounding sources. The 2 mm GISMO flux was nevertheless deboosted
to the best knowledge of the 2 mm number counts (Dwek, private communication). Using our simulations, we
find that the boosting factor for the AzTEC flux density is less than 5 per cent, which is smaller
than the calibration error, and hence insignificant.

Searching for possible optical counterparts in the CLASH catalogue, we find five sources within the
AzTEC beam (see Fig. \ref{mapaSN}), with optical photometric redshifts within the range of our estimated
(sub-)mm photometric redshift. We propose that one (or a blend) of these galaxies is the 
real counterpart of the AzTEC and GISMO sources.

We have also detected other AzTEC sources in our extended map with S/N$>$4 which also have {\it Herschel} counterparts.
These sources, however, are $>$30 arcsec from MACS1149-JD1. The analysis of these detections will be
presented in subsequent papers describing the AzTEC/LMT FF programme (Monta\~na et al. in preparation; 
Pope et al. in preparation).

\subsection{Constraints on the dust properties of MACS1149-JD1}

The AzTEC flux density upper limit of 0.51 mJy (3$\sigma$) at the position of MACS1149-JD1 can be used to obtain an upper limit on
the dust mass ($M_{\rm d}$) for this high-redshift galaxy, once  a dust temperature ($T_{\rm d}$) 
and a dust mass absorption coefficient are assumed. However, the strong dependence on $T_{\rm d}$ makes 
it difficult to obtain meaningful constraints on $M_{\rm d}$ unless reasonable temperature estimates are
available. In particular, for the upper limit of $M_{\rm d}$, it is more important to constrain the lower
limit of the dust temperature, since lower values of $T_{\rm d}$ result in higher dust masses 
(e.g. \citealt{2014MNRAS.443.1704H}). We compute the mass of dust by assuming a dust mass absorption
coefficient $\kappa_{\rm d}(\nu_{\rm rest})=0.15$ $(850 \mu\rm m/\lambda_{\rm rest})^\beta$ m$^2$ kg$^{-1}$
(\citealt{2003Natur.424..285D}), $\beta=1.5$  and by removing the contribution of the cosmic infrared background (CMB) to  dust 
heating as detailed by \citet{2013ApJ...766...13D}. Finally, to examine the range of $M_{\rm d}$, we assume
a dust temperature range of $T_{\rm d}=30-50$ K, where $T_{\rm d}$ is the dust temperature without the 
contribution due to the CMB heating. Using these values we estimate a dust mass limit of 
$M_{\rm d}<3.\times10^7 M_\odot/\mu - 2.4\times10^8 M_\odot/\mu$, where the lower limit corresponds to the
higher dust temperature. Correcting for the lens amplification ($\mu\approx15$; \citealt{2012Natur.489..406Z}),
the mass dust limit corresponds to $M_{\rm d}<2.1\times10^6 M_\odot - 1.6\times10^7M_\odot$.

Using the ${\rm SFR_{UV}}=1.2$ $M_\odot$ yr$^{-1}$ (\citealt{2012Natur.489..406Z}), the SFR-$M_{\rm d}$ 
relation found for local galaxies (e.g. \citealt{2010MNRAS.403.1894D}; \citealt{2012MNRAS.427..703S})
suggests an $M_{\rm d}\approx2\times10^7$ $M_\odot$. This value is  slightly higher than our upper limit, 
which may indicate a deficit of dust when compared to local galaxies. This has been predicted by 
\citet{2014ApJ...782L..23H} and has also been found for other high-z galaxies as Himiko at $z=6.6$ 
(\citealt{2013ApJ...778..102O}; see also \citealt{2014A&A...569A..98T}).

We also estimate the upper limit on the intrinsic infrared luminosity ($L_{\rm IR}$, $8-1000$ $\mu\rm m$) 
scaling different SED templates to our CMB-corrected $3\sigma$ AzTEC upper limit. We obtain 
$L_{\rm IR}<8\times10^{11} L_\odot/\mu - 1.2\times10^{12} L_\odot/\mu$. Correcting again for the lens 
amplification, this limit corresponds to $L_{\rm IR}<5\times10^{10} L_\odot - 8\times10^{10} L_\odot$. 
Using the \citet{1998ApJ...498..541K} relation, the total IR luminosity corrected for magnification 
converts to a dust-obscured star formation rate of ${\rm SFR_{\rm IR}}<9-14$ $M_\odot$ yr$^{-1}$.

A proxy for the UV attenuation ($A_{\rm{UV}}$) may be constrained by combining our observed limit on 
${\rm SFR_{\rm IR}}$ and the ${\rm SFR_{UV}}$ estimated by \citet{2012Natur.489..406Z}, where  
$A_{\rm{UV}}\approx2.5\log(\rm SFR_{\rm IR}/SFR_{UV})$, as in  \citet{2012ApJ...755...85M}. We obtained a limit of $A_{\rm{UV}}<2.1--2.7$ mag 
($A_{\rm{V}}<1.0-1.2$ mag, assuming the SMC attenuation curve). This limit is comparable to those derived 
for other Ly$\alpha$ and Lyman break galaxies at $z>6.5$ (\citealt{2015A&A...574A..19S};
\citealt{2015Natur.519..327W}).

\section{Summary}

We present $8.5$ arcsec FWHM angular resolution observations at 1.1 mm with the AzTEC camera on the LMT 
towards the $z\sim9.6$ galaxy MACS1149-JD1, for which a possible 2 mm detection with the GISMO 
camera has been reported. These new observations are a factor of $\sim3$ better in angular 
resolution and  a factor of $\sim4$  deeper than the GISMO observations assuming an average SMGs
template (\citealt{2010A&A...514A..67M}) at $z\sim1$ (or similar depth at $z\sim9.6$). However,
despite the achieved depth of $\sigma_{1.1mm}=0.17$ mJy in the AzTEC map, we have not found evidence of 1.1 mm 
continuum emission at the position of MACS1149-JD1.

A $3.5\sigma$ AzTEC detection at $11$ arcsec from MACS1149-JD1, consistent with the GISMO position, is the
most likely counterpart for the GISMO source. Combining the AzTEC and GISMO photometry with 
{\it Herschel} ancillary data we derive a $z_{\rm phot}= 0.7-1.6$, which further indicates that this
galaxy is not associated with MACS1149-JD1.

Finally, from the non-detection of the $z\sim9.6$ galaxy MACS1149-JD1, we derive the following 
($3\sigma$) upper limits corrected for gravitational lensing magnification and for CMB effects: 
dust mass $<1.6\times10^7$ $M_\odot$, IR luminosity $<8\times10^{10}$ $L_\odot$, 
SFR $<14$ $M_\odot$ yr$^{-1}$, and UV attenuation $<2.7$ mag. These values, which represent some of 
the highest redshift estimations for these quantities, are consistent with measurements in other 
$z>6.5$ galaxies.

\section*{Acknowledgements} 
This work would not have been possible without the longterm financial support from
the Mexican Science and Technology Funding Agency, CONACYT, during the construction and early
operational phase of the LMT, as well as support from the US NSF, the Instituto Nacional de 
Astrof\'isica, \'Optica y Electr\'onica (INAOE) and the University of Massachusetts (UMASS).
This work has been mainly supported by Mexican CONACyT research grants CB-2011-01-167291 and 
CB-2009-133260. JAZ acknowledges support from a CONACyT studentship. MJM acknowledges the 
support of the UK Science and Technology Facilities Council (STFC), British Council Researcher
Links Travel Grant, and the hospitality at INAOE. JSD acknowledges the support of the European
Research Council through the award of an ERC Advanced Grant. {\it Herschel} is an ESA space 
observatory with science instruments provided by European-led Principal Investigator consortia 
and with important participation from NASA. This work utilizes gravitational lensing models 
produced by PIs Brada{\v c}, Ebeling, Merten \& Zitrin, Sharon, and Williams funded as part
of the HST Frontier Fields programme conducted by STScI. STScI is operated by the Association of 
Universities for Research in Astronomy, Inc. under NASA contract NAS 5-26555. The lens models 
were obtained from the Mikulski Archive for Space Telescopes (MAST).

\bsp

\label{lastpage}

\end{document}